\documentclass{article}

\usepackage{PRIMEarxiv}

\usepackage[utf8]{inputenc} 
\usepackage[T1]{fontenc}    
\usepackage{hyperref}       
\usepackage{url}            
\usepackage{booktabs}       
\usepackage{amsfonts}       
\usepackage{nicefrac}       
\usepackage{microtype}      
\usepackage{lipsum}
\usepackage{fancyhdr}       
\usepackage{graphicx}       
\graphicspath{{media/}}     
\usepackage{siunitx}

\title{SPIM-Flow: An integrated light-sheet and microfluidics platform for Hydrodynamic Studies of \emph{Hydra}
\thanks{\textit{\underline{Citation}}: 
\textbf{Hedde PN, et al., SPIM-Flow: An Integrated light-sheet and fluidics platform for Hydrodynamic Studies of Hydra,  DOI:000000/11111.}}
}

\author
{Per Niklas Hedde,$^{1,\#}$ Erika L. Gomez,$^{2}$ 
Leora Duong,$^{3}$ Robert E. Steele,$^{4}$ Siavash Ahrar$^{2,5,\#}$\\
\\
\normalsize{$^{1}$Beckman Laser Institute and Medical Clinic, University of California Irvine, CA, USA}
\\
\normalsize{$^{2}$Department of Biomedical Engineering, CSU Long Beach, CA, USA}\\
\normalsize{$^{3}$ Department of Molecular Biology and Biochemistry, University of California Irvine, CA, USA}\\
\normalsize{$^{4}$ Department of Biological Chemistry, University of California Irvine, CA, USA}\\
\normalsize{$^{5}$ Department of Physics and Astronomy, University of California Irvine, CA, USA}\\
\\
\normalsize{\# authors contributed equally and to whom correspondence should be addressed}
}

\begin{document}
\maketitle

\newenvironment{sciabstract}{%
\begin{quote} }
{\end{quote}}

\begin{sciabstract}
Selective plane illumination microscopy (SPIM), or light sheet, is a powerful three-dimensional imaging approach. However, access to and interfacing microscopes with microfluidics have remained challenging. Complex interfacing with microfluidics has limited the SPIM's utility in studying the hydrodynamics of freely moving multicellular organisms. We developed SPIM-Flow, an inexpensive light sheet platform that enables easy integration with microfluidics. We used SPIM-Flow to study the hydrodynamics of a freely moving \emph{Hydra} polyp in millimeter-sized chambers (4 mm wide, 1.5 mm height). Our initial experiments across multiple animals, feeding on a chip (\emph{Artemia franciscana} nauplius used as food), and baseline behaviors (e.g., tentacle swaying, elongation, and bending) indicated animals' health inside the system. SPIM enabled easy imaging of the freely moving animal and tracer beads (for fluid visualizations) inside the larger chambers. Next, using the chambers, we investigated \emph{Hydra}'s response to flow. Results suggest that animals responded to established flow by bending and swaying their tentacles in the flow direction. Finally, we used a previously described video analysis software (FlowTrace) to visualize pathlines generated by (e.g., vortex generated by the tentacle sways) and around \emph{Hydra} (e.g., due to flow). These results demonstrated the SPIM-Flow's utility to study the hydrodynamics of freely moving animals. 
\end{sciabstract}

\newpage
\section{Introduction}

\subsection{Need for SPIM} Selective plane illumination microscopy (SPIM), or light sheet microscopy, is a powerful approach for three-dimensional (3D) imaging of biological samples at high spatio-temporal resolution\cite{Stelzer1}. Typically, the excitation arm (i.e., light sheet path) of the microscope is arranged at a 90-degree angle relative to the detection arm (i.e., camera path)\cite{Huisken1}. In the observation plane, the excitation light is restricted to a thin illumination plane via cylindrical optics or other beam shaping methods. This one dimensional confinement provides true optical sectioning, and limits phototoxicity and photobleaching of the sample, thus enabling long-term imaging \cite{Stelzer1}. A wide variety of  SPIM platforms have been demonstrated \cite{Olarte1}. Examples include open-source platforms \cite{Voigt1,Gualda1}, systems that use a single objective \cite{Yang1}, approaches that equip epifluorescence microscopes with light sheet illumination \cite{Yang1}, and even systems that use mobile phones for imaging \cite{Hedde1}. SPIM has enabled biological discoveries across sub-cellular imaging \cite{Hedde2, Sapoznik1}, developmental studies (across vertebrates, invertebrates, and plant model systems) \cite{Lemon1,silvestri1, von2017light}, and optically cleared tissues (e.g., entire brains of small animals) \cite{ueda2020whole}. While SPIM provides a powerful approach for long-term 3D imaging, samples are often housed in static conditions. Thus, interfacing SPIM with fluidics could further enhance the approach by allowing dynamic manipulation of a sample’s microenvironment \cite{zhou2022review}.

\subsection{Integrating SPIM with fluidics}
Various strategies have been explored to integrate fluidics with light sheet microscopy\cite{albert2019applications}. Such  integration has enabled cytometry platforms\cite{regmi2013light,lin2018label,fan2021microfluidic}, generation and visualization of droplets \cite{jiang2017droplet}, and optofluidic-based platforms for 3D visualizing of cells, \emph{C. elegans}, and \emph{Drosophila} embryos [21, 22, 23]. For example, a recent exciting study by Vanwalleghem et al. demonstrated the strength of integration for  studying larval zebrafish brain-wide activity while using flow as a stimulus \cite{vanwalleghem2020brain}. We have also explored using a light sheet configuration (based on an inverted epifluorescence microscope) which can accommodate chambers with flow \cite{hedde2017sidespim}. Despite this progress, integrating SPIM with fluidics has remained difficult. For example, the integrating requires modification to the microscope or significant modification to the microchambers \cite{albert2019applications}. To address this gap, we present SPIM-Flow, a simple and inexpensive system that readily integrates light sheet microscopy with fluidics. We used SPIM-Flow to investigate the hydrodynamic behavior of a freely moving \emph{Hydra} polyp in a millimeter-sized chamber (4 mm wide, 1.5 mm height).

\subsection{\emph{Hydra} as a model for biomechanics and hydrodynamics} 

\emph{Hydra} is a freshwater cnidarian studied for its remarkable regenerative abilities\cite{vogg2019model}. Its transparent, tube-shaped body is divided into three regions: head, body column, and foot. The head includes tentacles and the hypostome – a dome-like structure containing the mouth opening at its apex. \emph{Hydra} uses the foot (i.e., basal disc) to attach to surfaces. The shape and movement of the body are controlled via a hydrostatic skeleton where fluid pressure transmits forces \cite{vogg2019model,kier2012diversity}. With these characteristics, \emph{Hydra} is an excellent model system for studying biomechanics and hydrodynamics. A recent study, for example, using imaging and machine learning, showed that the behavioral repertoire of \emph{Hydra} can be divided into six fundamental components  (i.e., Elongation, Tentacle sway, Body sway, Bending, Contraction, and Somersaulting) \cite{han2018comprehensive}. In other studies, it was shown that \emph{Hydra} must tear a hole through its epithelial tissue to open its mouth \cite{carter2016dynamics,campbell1987structure}. Most of these experiments are conducted under static conditions. Using fluid chambers would allow one to expand such studies by modulating the microenvironments’ physical (e.g., flow) or chemical (e.g., transitory drug delivery) compositions.

In a recent study, Badhiwala et al. developed three microfluidic systems to study \emph{Hydra} \cite{badhiwala2018microfluidics}. These included a chamber to constrain the body column and enable electrophysiology, a perfusion chamber for constrained locomotion, and a quasi-2D plane behavioral chamber (200-600  µm heights). These investigators recently used a double-layer microfluidic system to mechanically stimulate \emph{Hydra} via pneumatic valves and measure neuronal responses via calcium imaging \cite{badhiwala2021multiple}. In particular, these platforms enabled neuronal studies that required  animal immobilization or small chambers. However, these restrictions limit the movement of the organism and may negatively impact its health. In our study, by taking advantage of SPIM-Flow, we sought to investigate \emph{Hydra}’s response to flow without dramatically restricting movement. We constructed chambers 4 mm wide and 1.5 mm tall. We used fluidics to modulate the hydrodynamic environment, for example introducing flow or prey. We used our light sheet system to visualize the hydrodynamics of animal movement. Our studies suggest that \emph{Hydra} typically remains in an elongated or swaying state inside the chambers without flow. \emph{Hydra} responds to flow initiation via contraction or tentacle swaying. Moreover, \emph{Hydra} typically reorients by body bending (head and tentacles) in the flow direction.

\section{Methods}
\subsection{Optical system} 

The excitation arm included a compact laser diode, an adjustable iris, and a cylindrical lens. To create the light sheet, similar to our prior efforts \cite{Hedde2}, we cut the central part of the elliptical beam of a laser diode (wavelength 445 nm or 488 nm) with the iris (Thor labs SM1D12), followed by a cylindrical lens (f=12 mm) that focused the light along one dimension to form a sheet of light at the observation plane. These components were housed inside a custom-designed 3D-printed enclosure. We used a cannon-like design (i.e., light-cannon) to bring the cylindrical lens close to the fluidic chip. The 3D-printed enclosure containing the optical components for excitation was then mounted on an XYZ micromanipulator. A light sheet of 7 µm thickness was created utilizing these components (see Supplemental for analysis). The excitation arm was mounted on an optical breadboard at a 90-degree angle relative to the detection arm. The detection arm included a custom-designed 3D-printed chip holder to bring the fluidic chip close to the light-cannon. Fluorescent signals were collected perpendicular to the excitation with an objective lens (10x, NA 0.25) followed by an emission filter (500 nm long pass) and a tube lens (f = 50 mm) for imaging with a CMOS camera (FLIR Blackfly S-U3-200S6C-C: 20 MP). SPIM-Flow’s system-level diagram is provided (Figure 1.A). SPIM-Flow was housed on a small optical breadboard (300 mm x 300 mm) and inside an opaque enclosure to shield the setup from external light. Micro-manager software was used to control the camera. Designs for the 3D parts will be available from a corresponding OSF page.

\subsection{Millifluidic system} 
We used laser cutting to build the molds used in the study instead of conventional lithographic approaches. The size of \emph{Hydra} (typically 0.5 -10 mm long) and our goal to investigate their unhindered hydrodynamics informed our choice. To this aim, channels were designed to be 4 mm wide and 5 cm long. We laser cut the design on acrylic sheets (1/16th inch, 1.5 mm thickness) via a desktop laser cutter. Plastic parts were permanently attached to a flat base (either a plastic part or a glass slide). The attachment prevented the unwanted PDMS accumulation under the mold during the silicone casting. After the mold fabrication, we used conventional protocols to cast silicone parts from the mold (1:10 linker to the base mass ratio followed by baking the molds on hotplates at 85°C until the PDMS were fully cured). Individual devices were cut from the mold using craft knives, and inlet/outlet holes were introduced using disposable (1 or 2 mm diameters) biopsy punches.

Next, each PDMS part was plasma bonded to a microscope glass slide via a Harrick Basic Plasma cleaner. We placed the edge of the PDMS part on the edge of the glass slide to minimize the distance between the channel and the light cannon (Figure 1.B and 1.C). Additionally, we took steps to remove any roughness from the outer edges of the devices to minimize potential imaging artifacts. Some roughness can be introduced to the outer edges of PDMS parts while cutting them via a craft knife. In a typical epifluorescent illumination, such roughness is of no concern,since the path of light is through the device top (typically PDMS) and glass bottom. In our system, however, the chamber is illuminated through a PDMS wall and viewed from the device top. We applied additional uncured PDMS to the device’s outer wall using a craft knife after plasma bonding and removed any roughness. We left the device (with the side with fresh PDMS facing up) at room temperature for three hours. This pause allowed the PDMS to spread evenly and partially cure. Then we backed the device at 95°C to ensure that the PDMS was entirely cured. During the early experiments, we also attached coverslips to the device roof. However, later experiments demonstrated that the coverslip did not significantly improve imaging.

\subsection{\emph{Hydra} culture}
All experiments were carried out using the PT1 transgenic line of \emph{Hydra vulgaris} \cite{steele2019reproductive}. PT1 contains two transgenes. One of the transgenes expresses green fluorescent protein under the control of the promoter from the gene encoding the Hym176B neuropeptide. The other transgene expresses the DsRed2 gene under the control of an actin gene promoter. This line expresses green fluorescent neurons and red fluorescent epithelial cells. PT1 was maintained in Hydra medium 4.0, which was prepared using house deionized water and contained 1 mM calcium chloride, 0.33 mM magnesium sulfate, 0.5 mM sodium bicarbonate, and 0.03 mM potassium chloride. The hydras were fed once a week with nauplii of \emph{Artemia franciscana} from San Francisco Bay (Brine Shrimp Direct, Ogden, UT). \emph{Hydra} cultures were kept in an incubator at 18°C on a 12 hour light/12 hour dark cycle.

\section{Results}
\subsection{Loading the \emph{Hydra}} We developed a simple protocol to minimize potential damage and increase the successful loading of animals. First, a device was partially (1/2) filled with media. We used a glass Pasteur pipette to retrieve an animal from a culture tube. Our biggest challenge was \emph{Hydra} adhering to the glass transfer pipette since \emph{Hydra} can rapidly adhere and remain firmly attached to surfaces. It is essential to transfer a \emph{Hydra} quickly. Therefore, we released the animals directly inside a device inlet or on top of the inlet with excess media from the pipette to a device. Next, we used liquid withdrawal with a micropipette (1 mL) or a syringe to draw the animal with the additional liquid inside the chamber. We sought to position each \emph{Hydra} toward the chamber’s center for ease of  imaging. To enable flow visualization, we added red fluorescent beads (1 µm diameter, Fluoro-Max Polymer microspheres) to the medium. We note that beads are also visible in GFP emission wavelength. 

After loading, the chamber was placed and secured on the 3D-printed chip holder. \emph{Hydra} medium with beads was delivered to the chamber using an external syringe pump (KDS Legato, 210P series) via 3 or 10 mL syringes. Typical of all microfluidics experiments, care was taken to ensure there were no air bubbles upstream of the device. However, small bubbles were not a problem. For example, early in the studies, we accidentally introduced air bubbles into  two chambers. In each case, the animal rapidly responded to the air bubble and the pressure by contracting/bending. Animals were able to survive these exposures. However, we chose to replace these animals with fresh ones.

At the start of the study, we sought to observe feeding behavior of \emph{Hydra} in the system to examine health. After loading a \emph{Hydra} on a chip, multiple prey animals (\emph{Artemia franciscana}) were added to the chamber. We first used the imaging without its light sheet capabilities to independently investigate the potential chamber effects. \emph{Hydra} could successfully capture and eat multiple (2 to 3) \emph{Artemia} inside the chamber. Next, we verified similar feeding behavior using the SPIM imaging capabilities across independent animals (Supplemental Figure 1, and Supplemental Video 1) - video and images are from two different animals. After feeding, \emph{Hydra} ignored the remaining prey in the chamber. The feeding results (across N=4 different animals) suggested that the fluid chambers and the light sheet did not negatively impact the organisms' health. Experiments presented in the study took 2-3 hours; however, we have confirmed that a \emph{Hydra} can be kept alive on the same chip for multiple days.

\subsection{Static conditions and baseline behaviors}
Next, we examined the animals' health and sought to demonstrate the feasibility of observing various behaviors via SPIM-Flow without flow. Given the device's dimension, the animals' overall movements were unhindered. However, some animals could stretch across the entire 4 mm width of the chamber. Using the light sheet, we illuminated different regions of the freely moving animals. \emph{Hydra} responds to light as a stimulus \cite{trembley1744memoires,wilson1891heliotropism,haug1933lichtreaktionen,singer1963photodynamic}.To minimize the role of light as a sudden stimulus, we started recording an experiment 5 minutes after the light introduction. A typical recording session would last between 2-3 hours. SPIM's capability to limit the light exposure (only to the illuminated sheet) prevented photobleaching or phototoxicity. However, given the role of light as a stimulus, additional studies are required, and  SPIM-Flow could provide a versatile tool for this aim. We observed that most animals typically adhered to the surfaces (PDMS wall or the glass slide) closer to the direction of excitation light (wavelength 485 nm or 455 nm). Without flow, we observed behaviors ranging from elongation, and tentacle movement/swaying. (Figures 2 and 3, Supplemental Figure 2, Supplemental Video 3). Most animals would  remain in the chambers elongated or in a tentacle swaying state. Despite the chambers' large size, we did not observe somersaulting behaviors. These observations further supported animals' overall health and helped demonstrate that various repertoires of animal movements can be captured via SPIM-Flow.

\subsection{Response to flow} 
Next, we investigated \emph{Hydra} response to flow. \emph{Hydra} lives in bodies of freshwater that may experience local flow patterns due to environmental perturbations or currents in streams and rivers. Therefore, tools to investigate their hydrodynamic response to flow could provide valuable insights into the organisms' biomechanical and hydrodynamic lifestyle. In their study, Badhiwala et al.  \cite{badhiwala2018microfluidics} commented that animals typically bend in the flow direction. We sought to investigate this observation by utilizing our larger channels (1.5 mm in height, 4 mm wide). Similar to the Badhiwala et al. study, our results (across four independent animals) suggest that \emph{Hydra} typically responds to flow by bending and swaying their tentacles in the flow direction. Reversing the direction of flow resulted in \emph{Hydra} redirecting accordingly (Figure 4). For example, in results from Figure 4, the animal responded to the volumetric flow rate of 2 mL/hr.  Larger animals could withstand flow rates as large as 50 mL/hr. Our results also suggest that an animal could immediately respond to the flow initiation via rapid contraction or tentacles swaying in the direction of the flow (Supplemental Figure 3). However, additional experiments are required to better understand the immediate response to flow initiation. Utilizing SPIM-Flow, we briefly explored a proof of concept experiment, to investigate \emph{Hydra}’s response to pulsatile flow. Our preliminary results suggest that pulsatile flow resulted in an animal stretching towards the direction of the flow. 

Finally, we used a previously described flow visualization algorithm, FlowTrace, to help better visualize the hydrodynamic patterns observed via SPIM-Flow.  FlowTrace is a powerful yet simple algorithm that enables the extraction of flow features from video recordings \cite{gilpin2017flowtrace,gilpin2017dynamic}. Unlike typical pipelines for particle image velocimetry, FlowTrace is not computationally taxing. FlowTrace has been used to extract pathlines to visualize the time-varying flow fields generated by starfish larvae and other organisms. We used FlowTrace analysis of the SPIM-Flow videos to visualize flow fields (e.g., vortex generated by the tentacle sways) and pathlines generated by and around \emph{Hydra}. For this, we used the algorithm as a FIJI plugin \cite{gilpin2017flowtrace}. Figure 5 demonstrates the original recording (A), and the corresponding FlowTrace visualizations (B). Pathline visualizations corresponding to Figure 4 and FlowTrace videos are provided as part of the supplemental material.

\section{Discussion and Conclusion}

Despite tremendous progress, access to SPIM imaging and interfacing the conventional systems with modern fluidics have remained challenging. We sought to address this gap by establishing SPIM-Flow. Our platform is an inexpensive light sheet microscope readily designed for integration with microfluidics. To validate the usefulness of SPIM-Flow, we used the platform to study the hydrodynamics of freely moving \emph{Hydra} in static and dynamic conditions. SPIM-Flow could enable various applications; for example, we explored its use as an image-based flow cytometer. However, as demonstrated, it could serve as a powerful tool to visualize and manipulate whole organisms and their microenvironment. Light sheet imaging specifically enabled easy visualization of tracer beads to extract hydrodynamic features. As a comparison, we also imaged the animals inside the chambers via a standard fluorescent microscope (Nikon Eclipse Ti-E microscope 10X objective - Supplemental Figure 5). Simultaneously visualizing the beads and the animal (due to the high signal from the animal and the background volume) were difficult with a standard microscope. Indeed SPIM-Flow is not a replacement for a standard fluorescent microscope. Instead, it is a complementary and inexpensive tool which enables hydrodynamic studies of organisms, such as \emph{Hydra}. Our initial experiments (feeding and observing various behaviors) supported that the platform did not impact animals' health. Next, we used microfluidics to investigate Hydra's response to continuous flow. Our results across multiple animals (N=4) suggest \emph{Hydra} responds to flow by bending in the flow direction. Reversing the direction of flow led to animals' reorientation. Utilizing the light sheet imaging, we visualized pathlines (via fluorescent beads) throughout the study. Moreover, using the FlowTrace algorithm, we produced visualizations highlighting the complex hydrodynamics of an animal movement. 

Our study demonstrated the platform's capability for studying freely-moving model organisms. In the future, we seek to explore the biomechanics of \emph{Hydra} movement further (e.g., quantifying displacements generated by the animals or systematically exploring animals' response to pulsatile flow). Microfluidics combined with conventional imaging has provided excellent platforms to study model systems such as zebrafish, \emph{Drosophila}  embryos, and \emph{C. elegans} \cite{kim2018microfluidics,hwang2013microfluidic}. Other emerging model systems could (and indeed have started to) benefit from need-driven systematic integration of micro and millifluidics and novel imaging techniques \cite{krishnamurthy2020scale,hol2020biteoscope,kumar2021microfluidic}. SPIM-Flow, as a simple and inexpensive  platform, can contribute to these studies with a focus on hydrodynamics of small animals.

\newpage
\section* {Acknowledgments}  
The authors express gratitude to Prof. Albert Siryaporn for his support.\\ 
This work was supported in part by CSULB startup funds and a CSUPRB research grant to SA. \\Additionally, this work was supported by National Institutes of General Medical Sciences R21GM135493 to PNH. \\LD is supported by an NSF GRFP (base number 2021317020).

\section*{Corresponding Authors Addresses}

Siavash Ahrar (Ph.D.)\\
Mail: Department of Biomedical Engineering, CSU Long Beach, 
1250 Bellflower Blvd. \\VEC-404.A, Long Beach, CA 90840, 
E-mail: siavash.ahrar@csulb.edu

Per Niklas Hedde (Ph.D.)\\
E-mail: phedde@uci.edu

\section*{Affiliations}

\begin{itemize}
\item PNH: Beckman Laser Institute and Medical Clinic, University of California Irvine, CA, US.
\item ELG: Department of Biomedical Engineering. CSU Long Beach, CA, US.
\item LD: Department of Molecular Biology and Biochemistry,University of California Irvine, CA, US.
\item RES: Department of Biological Chemistry, University of California Irvine, CA, US.
\item SA: Department of Biomedical Engineering. CSU Long Beach, CA, US.\\
          Department of Physics and and Astronomy, University of California Irvine, CA, US.
\end{itemize}

\newpage
\textbf {List of Figures}:
\begin{itemize}
  
  \item Figure 1: SPIM-Flow system diagram.
  \item Figure 2: Baseline behaviors during static conditions as an indicator for overall health.
  \item Figure 3: Baseline behaviors, bending and tentacle movement, during static conditions.
  \item Figure 4: Response to flow. 
  \item Figure 5: Hydrodynamic visualization via FlowTrace. 
\end{itemize}

\textbf {List of Supplementary Figures}:
\begin{itemize}
  
  \item Sup Figure 1: \emph{Hydra} feeding behavior as an indicator for health.
  \item Sup Figure 2: \emph{Hydra} baseline behavior, elongation, during static conditions as an indicator for overall health.
  \item Sup Figure 3: \emph{Hydra} responses to the initiation of flow. 
  \item Sup Figure 4: FlowTrace visualizations - response to flow.
  \item Sup Figure 5: Imaging \emph{Hydra} via a standard fluorescent microscope.
  
\end{itemize}

\textbf {List of Supplementary Videos}:
\begin{itemize}
  
  \item Movie S1: \emph{Hydra} feeding behavior 
  \item Movie S2: \emph{Hydra} tentacle sway behavior 
  \item Movie S3: \emph{Hydra} response to flow
  \item Movie S4: FlowTrace - example 1 (vortex)
  \item Movie S5: FlowTrace - response to flow example 2
\end{itemize}

\newpage
\begin{flushleft}
\bibliographystyle{unsrt}
\bibliography{refs} 
\end{flushleft}

\newpage
\section*{Supplemental section - analysis of light sheet properties}

The Rayleigh criterion defines the optical resolution, \emph{r}, as:

\begin{center}
\emph{r} = 0.61 $\frac{\lambda}{NA}$
\end{center}

where $\lambda$ is the wavelength and NA the numerical aperture of the lens. 

For small aperture lenses, the NA can be approximated by:
\begin{center}
NA = $n \frac{D}{2f}$
\end{center}
where D is the lens diameter, \emph{f}the lens focal length, and n the refractive index of the surrounding medium, which in our case was air (n = 1). 

Combining the two equations yields:

\begin{center}
\emph{r} = 1.22 $\frac{\lambda f}{D}$
\end{center}

With an excitation wavelength of 488 nm, a beam diameter of 1 mm, and the focal length of the cylindrical lens of 12 mm, the light sheet minimum thickness was 7.1 µm.

\newpage

\begin{figure}[b]
\includegraphics[width=\textwidth]{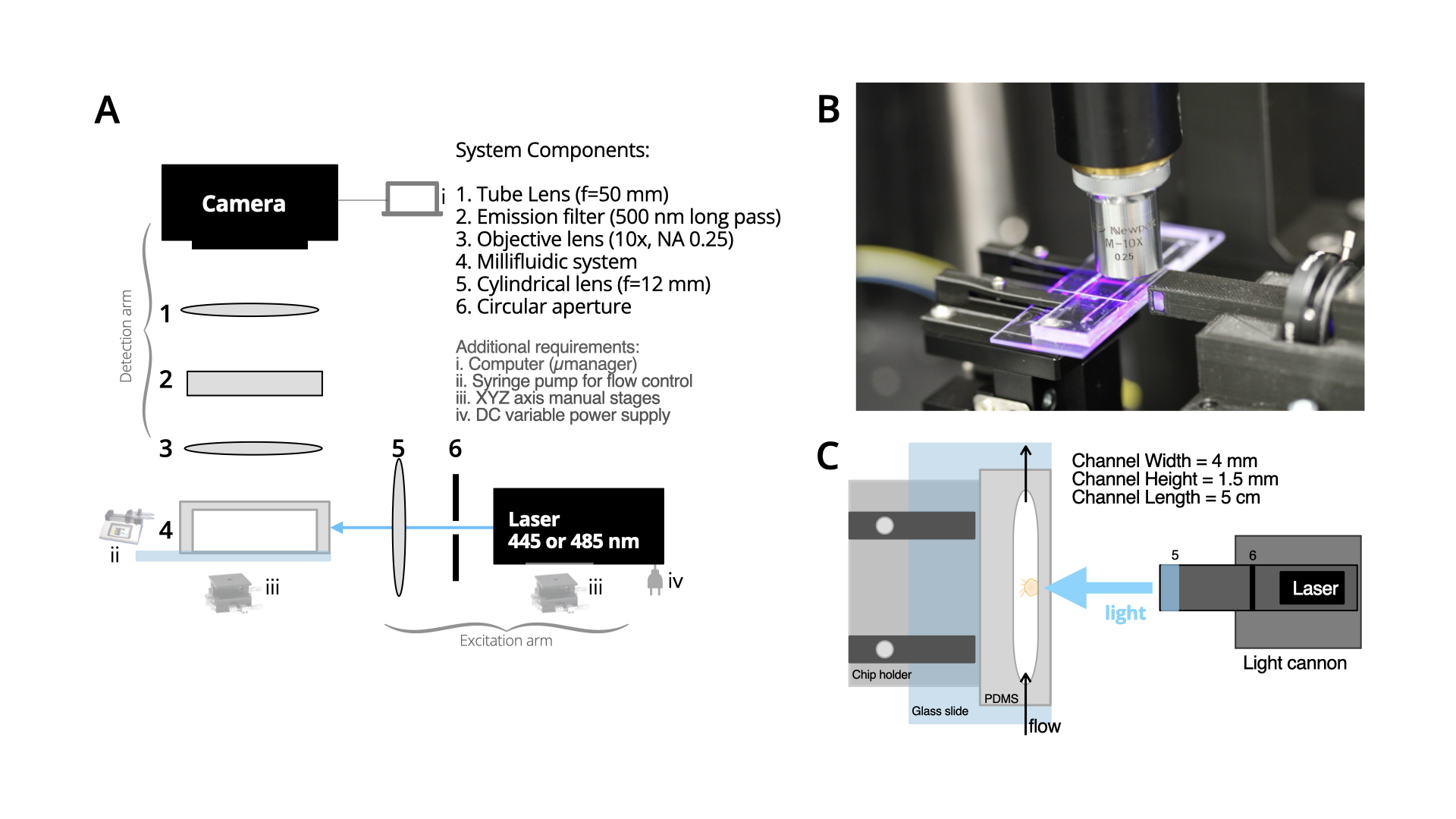}
\centering
\caption{\textbf {SPIM-Flow system diagram}. (A) Simple and inexpensive SPIM designed for compatibility with fluidics. (B) Photograph of the excitation arm (i.e., light cannon), microfluidic chamber held on the 3D printed stage, and the objective lens collecting fluorescent signals from the chamber. (C) Diagram of the PDMS chamber with its critical dimensions, stage and the Light cannon. Steps were taken to remove any roughness from the side of the chamber that accepts the light and to remove any rough edges that could create imaging artifacts. }
\end{figure}

\begin{figure}[b]
\includegraphics[width=\textwidth]{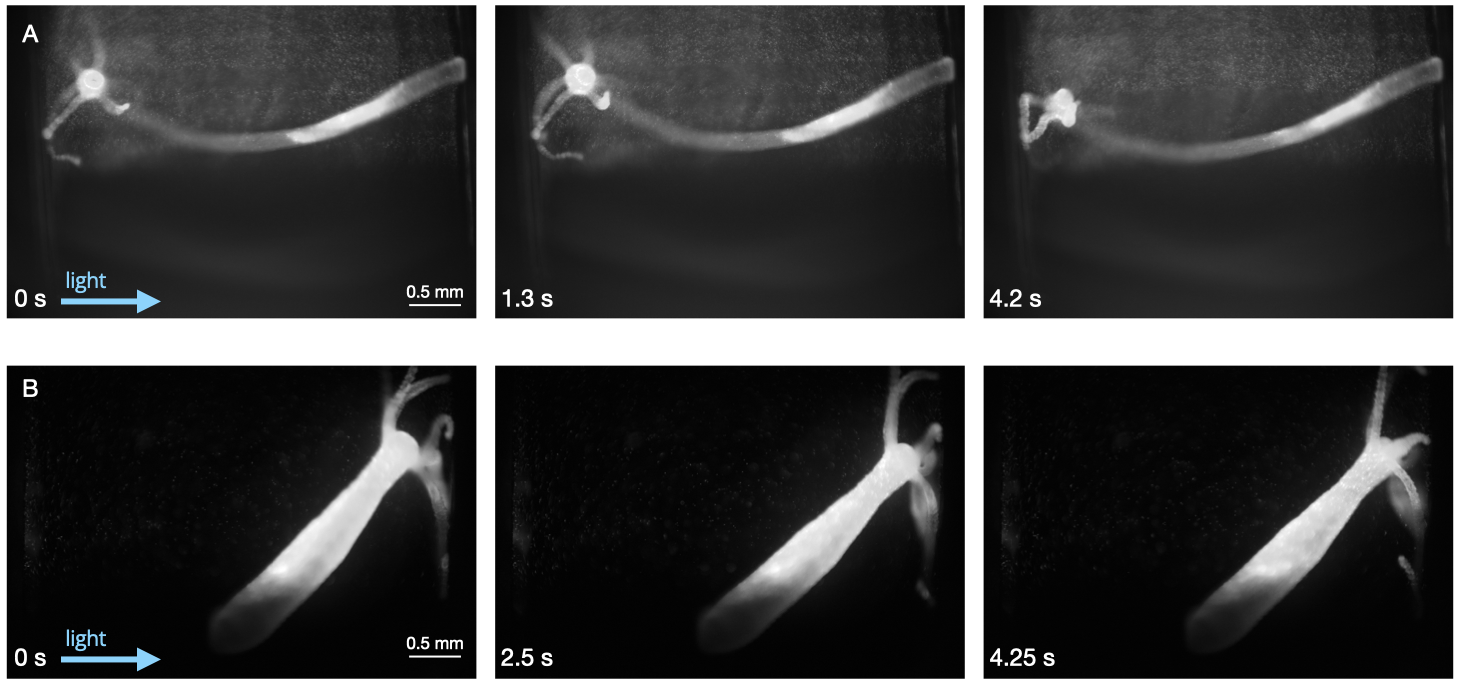}
\centering
\caption{\textbf {Baseline behavior during static conditions as an indicator for overall health}. (A) Elongation and tentacle movement. The  head (hypostome and tentacle ring)  are also visible. (B) A different animal in an elongated state inside a chamber. Using SPIM-Flow, GFP-positive  neurons can be visualized depending on the orientation and state of the \emph{Hydra}.}
\end{figure}

\begin{figure}[b]
\includegraphics[width=\textwidth]{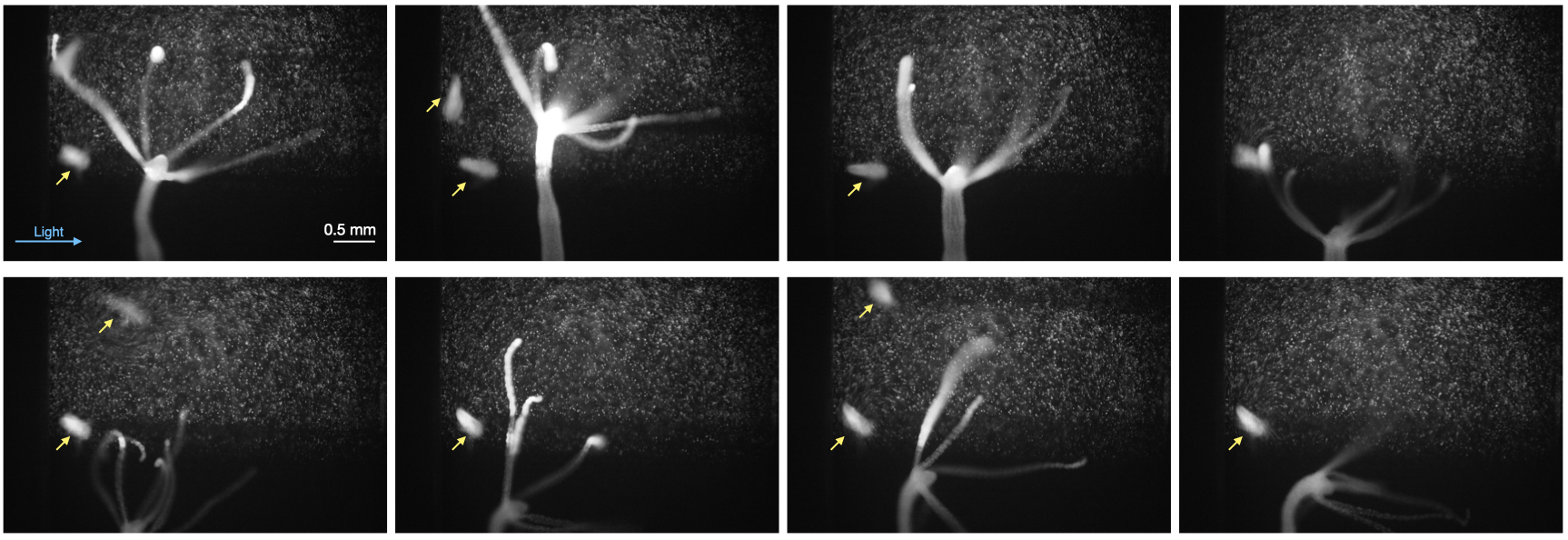}
\centering
\caption{\textbf {Baseline behaviors, bending and tentacle movement, during static conditions}.  Yellow arrows point to \emph{Artemia franciscana} nauplii that were ignored by the  \emph{Hydra} after capturing 3 nauplii earlier.}
\end{figure}

\begin{figure}[b]
\includegraphics[width=\textwidth]{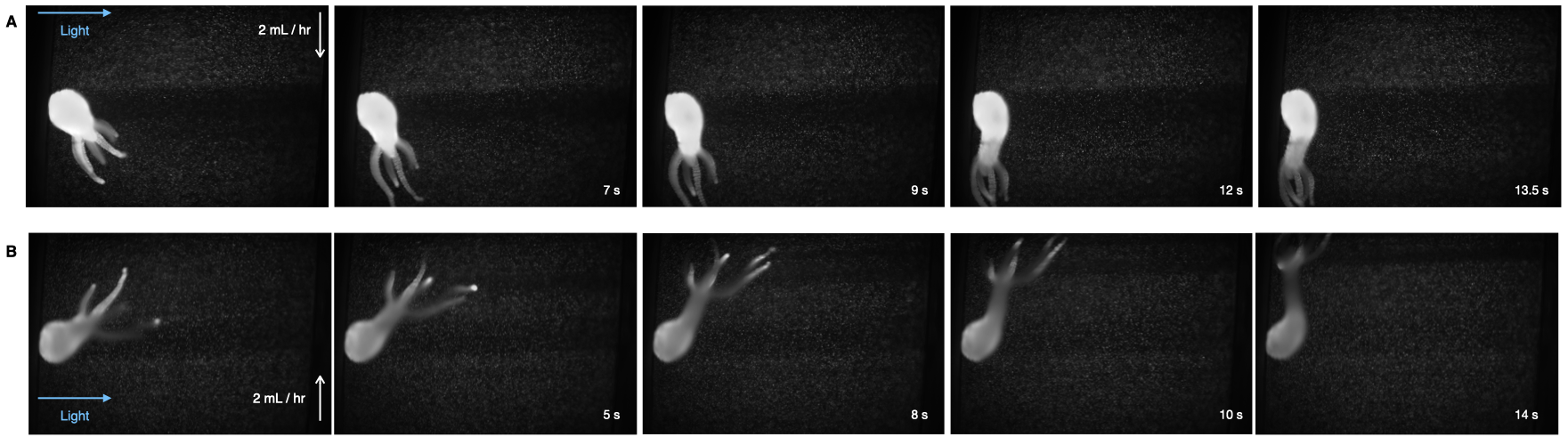}
\centering
\caption{\textbf {Response to Flow}.  Our results suggest that \emph{Hydra} typically bends and sways its tentacles in the direction of the flow. (A) Initiation of the flow and redirection of the animal towards the flow. White arrow demonstrates the flow direction (B) Reversing the flow direction, led to the animal redirecting with the new flow direction. }
\end{figure}

\begin{figure}[b]
\includegraphics[width=\textwidth]{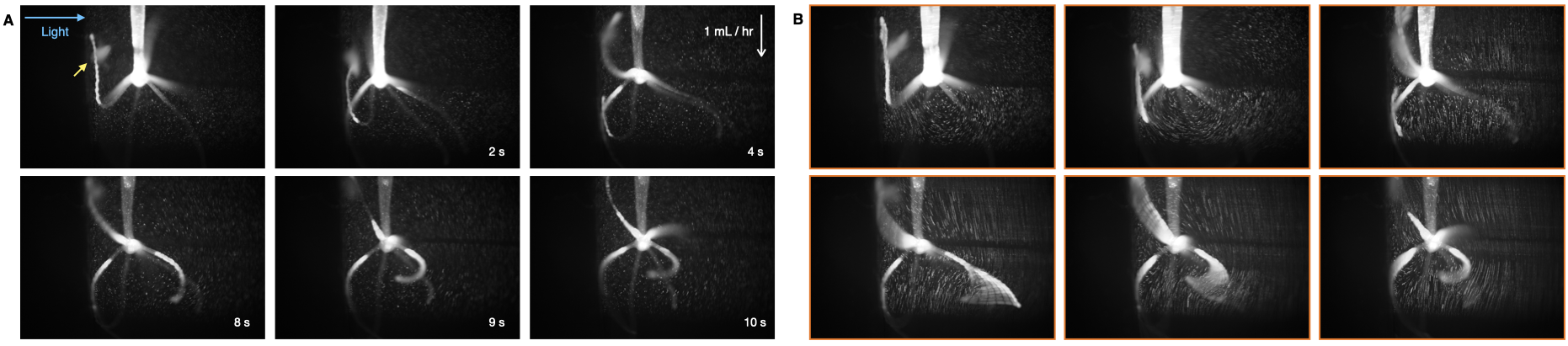}
\centering
\caption{\textbf {Hydrodynamic visualization via FlowTrace}. (A) Original recording via SPIM-Flow.  Flow starts at 4s. (B) Extracting the pathline to visualize the flow field generated by and around \emph{Hydra} via the FlowTrace algorithm.}
\end{figure}

\begin{figure}[b]
\includegraphics[width=\textwidth]{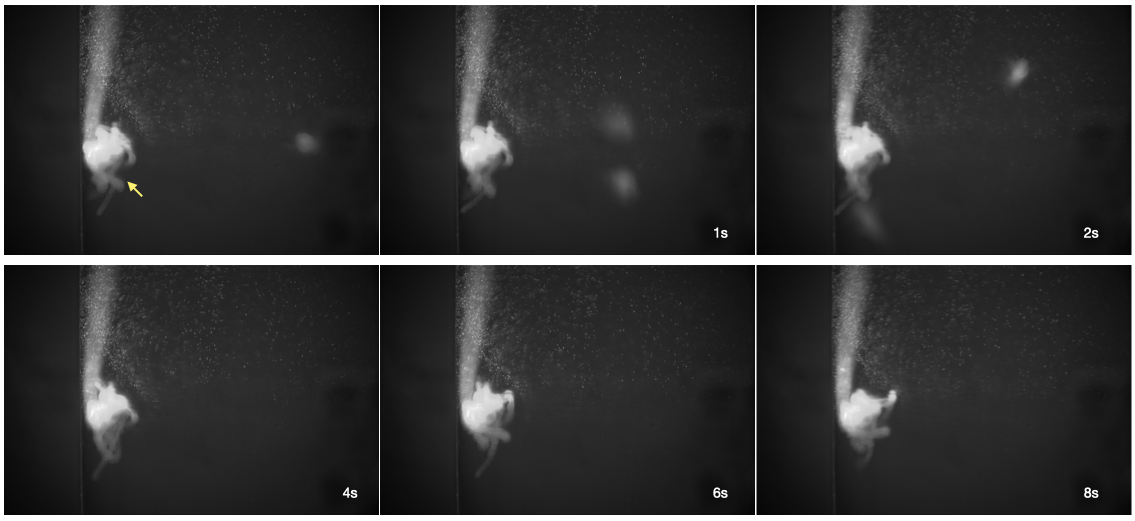}
\centering
\caption{\textbf {Supplemental Figure 1 \emph{Hydra}'s feeding behavior as an indicator for health}. Arrow indicates tentacles holding a \emph{Artemia franciscana} nauplius (brine shrimp). }
\end{figure}

\begin{figure}[b]
\includegraphics[width=\textwidth]{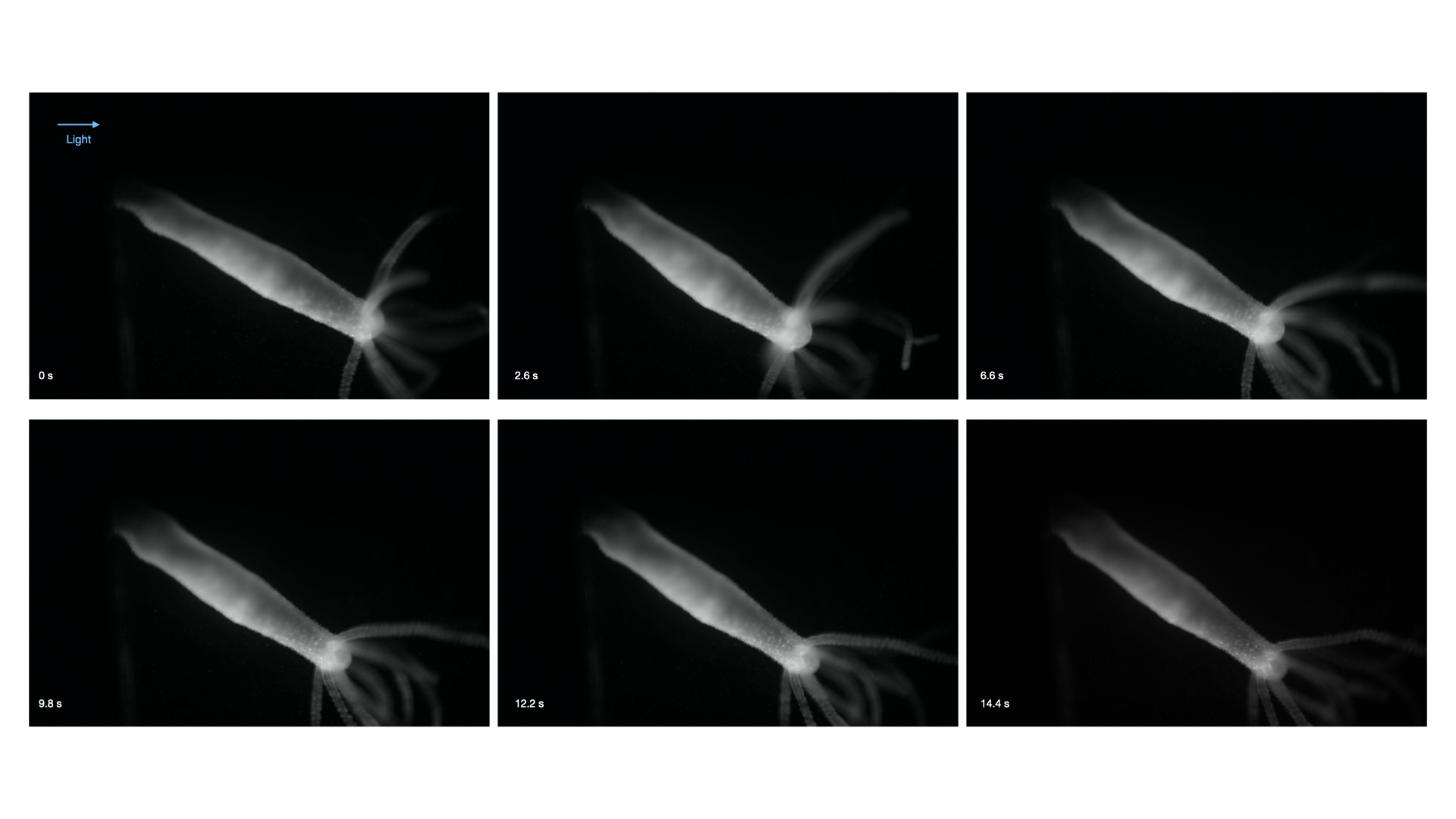}
\centering
\caption{\textbf {Supplemental Figure 2 \emph{Hydra}'s baseline behavior in static conditions}. Elongation state as an indicator for overall health}
\end{figure}

\begin{figure}[b]
\includegraphics[width=\textwidth]{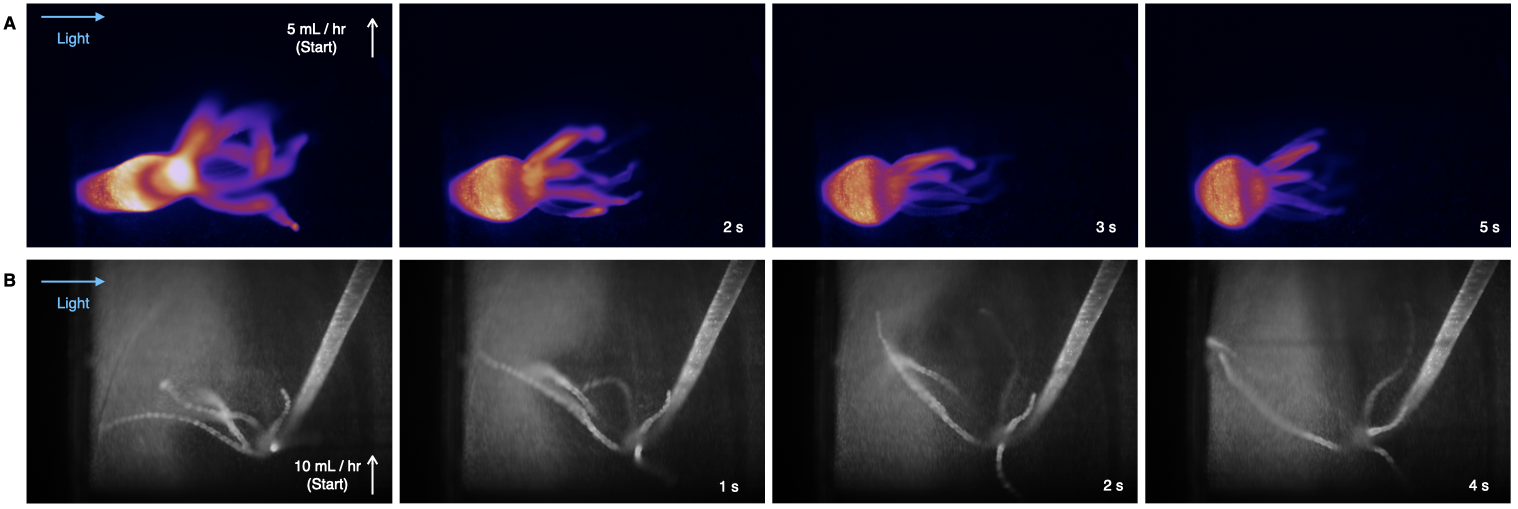}
\centering
\caption{\textbf {Supplemental Figure 3 Responses to the initiation of flow}.
(A) A \emph{Hydra} contracting in response to the flow initiation. (B) A \emph{Hydra} responding to the flow initiation by tentacle swaying behavior.  Additional experiments are required to investigate the dynamic response of \emph{Hydra} to various mechanical stimuli.}
\end{figure}

\begin{figure}[b]
\includegraphics[width=\textwidth]{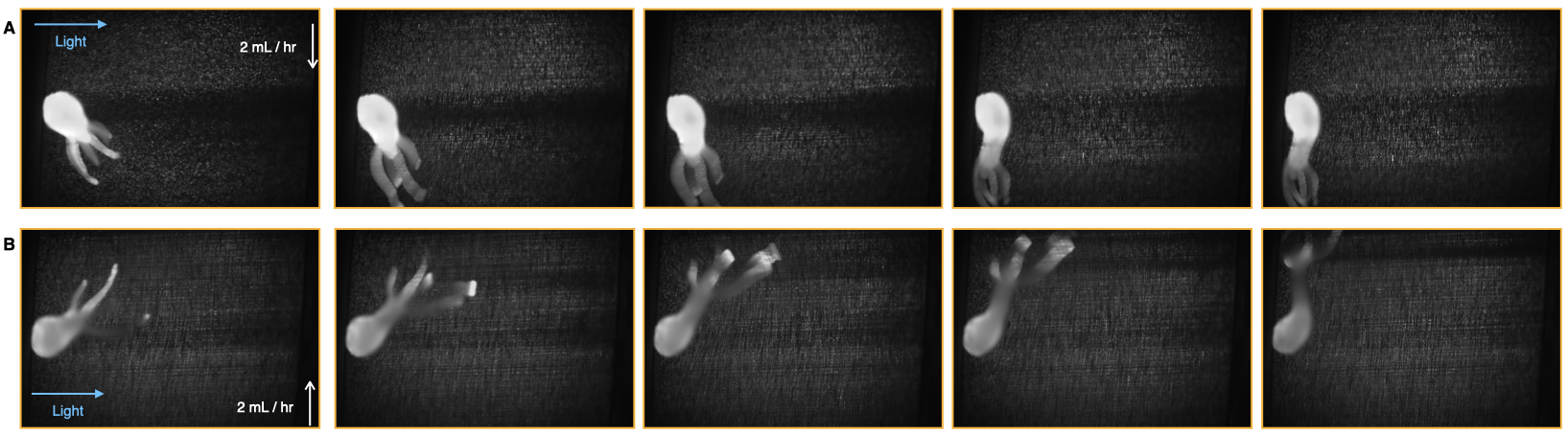}
\centering
\caption{\textbf {Supplemental Figure 4 FlowTrace visualizations}. Pathlines corresponding to \emph{Hydra}’s response to established flow.  Data corresponds to results from Figure 4 from the main text.}
\end{figure}

\begin{figure}[b]
\includegraphics[width=\textwidth]{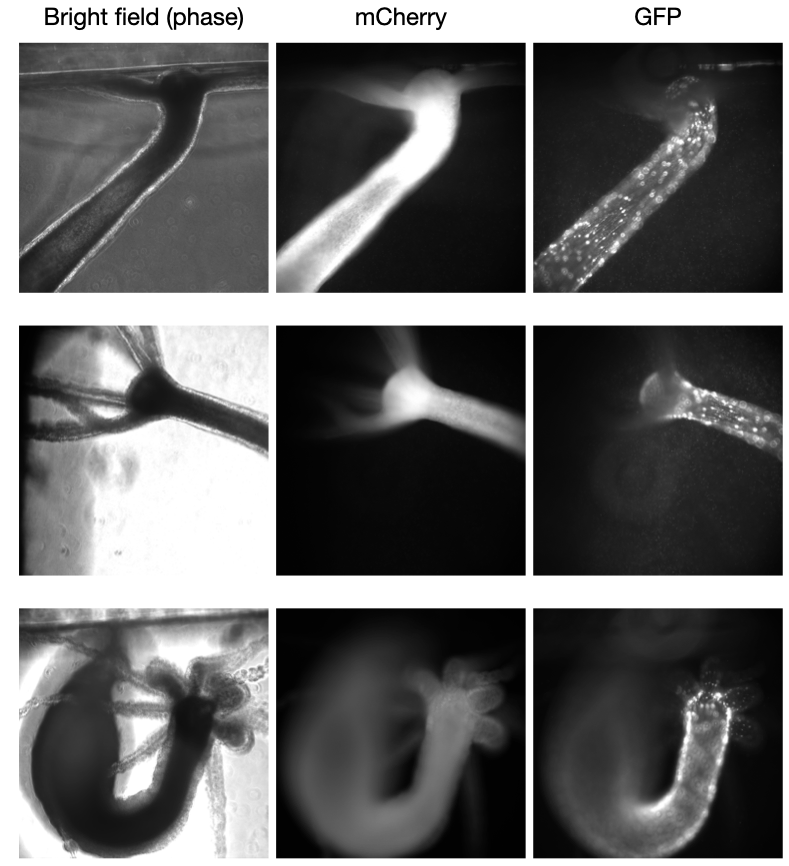}
\centering
\caption{\textbf {Supplemental Figure 5 Imaging \emph{Hydra} (N=2) via a standard fluorescent microscope}. Uisng a commercialfluorescent microscope (Nikon Eclipse Ti-E microscope 10X objective) - with a Sola light engine (Lumencor, Beaverton, OR), and 474/27 nm and 575/25 nm filters for excitation and 525/45 nm and 641/75 nm filters for emission for the GFP and DsRed2 visualizations). More anatomical features can be visualized with the standard microscope (. However, SPIM-Flow provides a frugal and simple alternative for hydrodynamic studies. Moreover, simultaneous visualization of the animal and the beads with the standard microscope would require additional optimization.}
\end{figure}

\end{document}